\documentclass[twocolumn,preprintnumbers,amsmath,amssymb]{revtex4}

\usepackage{graphicx}
\usepackage{dcolumn}
\usepackage{bm}
\usepackage{textcomp}
\begin{document}

\preprint{L03-0901}

\title{Anomalous Phase Transition in Strained SrTiO$_3$ Thin Films}

\author{Feizhou He}
\author{B. O. Wells}%
\affiliation{Department of Physics, University of Connecticut,
Storrs, CT 06269}%

\author{S. M. Shapiro}
\author{M. v. Zimmermann}
 \altaffiliation[Present Address: ]{HASYLAB at DESY, Notkestr.
 85, 22603 Hamburg, Germany}
\affiliation{Department of Physics, Brookhaven National Lab,
Upton, NY 11973}%

\author{A. Clark}
\author{X. X. Xi}
\affiliation{Department of Physics, The
Pennsylvania State University, University Park, PA 16802}

\date{\today}

\begin{abstract}
We have studied the cubic to tetragonal phase transition in
epitaxial SrTiO$_3$ films under various biaxial strain conditions
using synchrotron X-ray diffraction. Measuring the superlattice
peak associated with TiO$_6$ octahedra rotation in the low
temperature tetragonal phase indicates the presence of a phase
transition whose critical temperature is a strong function of
strain, with T$_C$ as much as 50K above the corresponding bulk
temperature. Surprisingly, the lattice constants evolve smoothly
through the transition with no indication of a phase change. This
signals an important change in the nature of the phase transition
due to the epitaxy strain and substrate clamping effect. The
internal degrees of freedom (TiO$_6$ rotations) have become
uncoupled from the overall lattice shape.
\end{abstract}

\maketitle

Perovskite films have received a great deal of interest lately
due to the potential for creating working technologies based on a
variety of interesting properties such as high-T$_C$
superconductivity, colossal magneto-resistivity,
ferro-electricity, and variable dielectric constants. These
properties can be quite different in thin films versus nominally
similar bulk samples. The primary reasons for the changes in
properties are believed to be strain and defects. There is also
an emerging theoretical effort to treat the effects of strain. To
understand the mechanisms for these changes we must be able to
conduct detailed microscopic measurements of the atomic and
electronic structure.

SrTiO$_3$ (STO) is a nearly ferroelectric material with a large
dielectric nonlinearity and low dielectric loss at cryogenic
temperature, making it ideal for tunable microwave
devices\cite{Xi00}. STO is also a good model system for studying
structural phase transitions. Bulk STO crystals are cubic at room
temperature, with space group Pm3m(O$_h^1$), but become
tetragonal, space group I4/mcm(D$_{4h}^{18}$), below about 105K.
This phase transition involves the rotation of TiO$_6$ octahedra
and has been featured historically in the study of structural
phase transitions as the classic example of a soft-mode phase
transition\cite{Uno67}. The unit cell of the tetragonal phase has
a volume four times that of the cubic unit cell, with approximate
unit cell of
$\sqrt{2}$\textit{a}$\times$$\sqrt{2}$\textit{a}$\times$2\textit{a},
where \textit{a} is the lattice parameter of cubic unit cell. In
this paper we describe the tetragonal phase as pseudo-cubic, in
order to compare the structure before and after the phase
transition. In this pseudo-cubic frame, the tetragonal phase has
additional superlattice peaks at half integer index positions.
These superlattice peaks disappear as the temperature is raised
through the tetragonal-cubic phase transition.

In the bulk, the deviation from cubic symmetry is directly related
to the rotation angle of the TiO$_6$ octahedra. This rotation
angle has been identified as the order parameter for this phase
transition\cite{Mul68}. Correspondingly, the intensities of the
superlattice peaks are proportional to the square of the order
parameter and can be used to track the phase transition.

For a second-order structural phase transition, usually the volume
of the unit cell changes smoothly through the transition
temperature, but lattice parameter versus temperature curves have
a sudden change in slope. This is the case for bulk
STO\cite{Lyt64}\cite{Ale69}. For epitaxial films the in-plane
lattice parameters are subject to lateral restraint from the
substrate and therefore do not have the freedom to change as in
bulk. In this paper we show that for epitaxial films of STO the
lattice parameters are indeed constrained though this does not
completely inhibit the occurrence of the soft-mode phase
transition. The strains do change both the transition temperature
and the nature of the transition.

The films studied were STO films grown on (001) LaAlO$_3$ (LAO)
single crystal substrate with SrRuO$_3$ (SRO) buffer layers. The
samples were grown using the Pulsed Laser Deposition technique
with details described elsewhere\cite{Li98a}. X-ray diffraction
implies excellent epitaxy with average mosaics around 0.2 degrees
and no detectable misoriented regions. Several different
measurements of films grown in the same system under the same
conditions have been reported elsewhere\cite{Sir99}\cite{Li98b}.
All of the samples were of the type STO/SRO/LAO with variations of
either the STO or SRO thickness. The main study consisted of STO
films with thickness varying from 50 to 1000nm on a SRO-buffer
layer 350nm thick. A secondary study investigated STO films 200nm
thick on SRO buffer layers with thickness from 0 to 350nm. X-ray
diffraction measurements were carried out at beamline X22A at the
National Synchrotron Lights Source at Brookhaven National
Laboratory. X22A has a bent Si(111) monochromator, giving a small
beam spot and fixed incident photon energy of 10 keV. The
longitudinal resolution with a Si(111) analyzer was at least
0.001\AA$^{-1}$ (HWHM) for an (0,0,2) peak, as measured from the
LAO substrate. Below room temperature the sample was cooled in a
closed cycle refrigerator with a temperature control of
$\pm$0.5\textdegree.

\begin{figure}
\includegraphics[scale=.8]{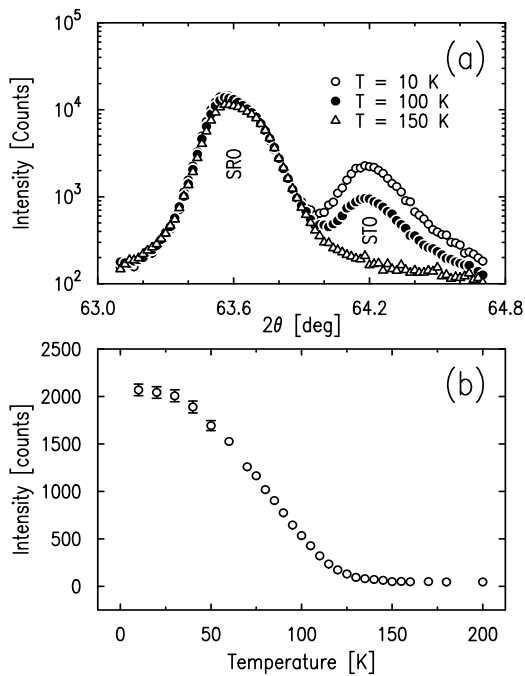}
\caption{\label{fig:fig1.eps}(a) (1/2. 1/2, 7/2) peaks of STO and
SRO at different temperature, (b) intensity of STO superlattice
peak versus temperature.}
\end{figure}

We choose the coordinate system such that the axis normal to the
film surface is the \textit{c} axis. Like bulk STO, all of our STO
films have extra half-integer superlattice peaks at low
temperature. The intensity of these peaks decreases as the
temperature rises and the peaks vanish at 130$\sim$170K for
different samples. Thus the STO films also have a phase transition
with similar symmetry change as seen in the bulk.
Fig.~\ref{fig:fig1.eps} shows an example of such a superlattice
peak for a [1000nm STO / 350nm SRO] sample. In the first panel are
raw data scans through the (1/2, 1/2, 7/2) peak at some
representative temperatures. The presence of a superlattice peak
for SRO is a result of its structural phase transition occuring at
$\sim$800K. The STO and SRO peaks are well resolved and the STO
peak intensity decreases with increasing temperature.
Fig.~\ref{fig:fig1.eps} panel (b) is a summary plot of the peak
intensity versus temperature. Compared to the bulk, the intensity
versus temperature curve is shifted towards higher temperature.
The intensity does not fall abruptly to zero but includes a tail
that extends to higher temperature. This indicates that the phase
transition is rounded and occurs over a range of temperature. This
rounded transition is most likely due to the presence of a range
of local strains within the film. Due to the rounding of the
transition, we need to derive a method for extracting the average
T$_C$. We extrapolate the linear portion of the curve between 10\%
to 40\% maximum intensity and define the transition temperature as
where this line crosses the zero intensity. The use of a linear
function is justified by the data itself as well as previous work
that has found a power law coefficient b very close to one for
this phase transition in bulk samples\cite{Doi00}. While this
method may produce T$_C$ too high by a few degrees ($<$5K), it
allows for an accurate comparison of T$_C$ between different
films. We believe we can determine $\Delta$T$_C$ to about 0.5K.

\begin{figure}
\includegraphics[scale=.8]{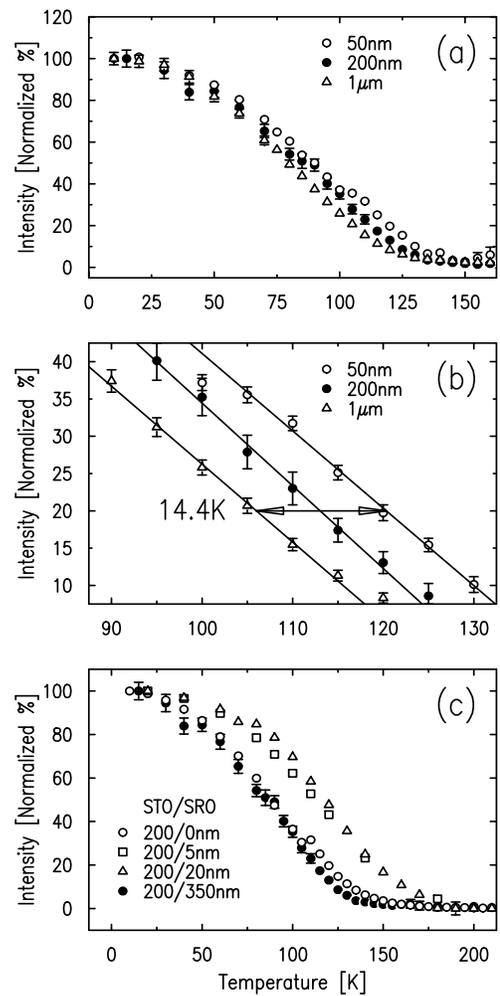}
\caption{\label{fig:fig2.eps}(a) Temperature dependence of peak
intensity for STO films of different thickness, (b) detail of (a),
shows the change of temperature of phase transition,
$\Delta$T$_C$, (c) temperature dependence of peak intensity for
STO films on different buffer layers.}
\end{figure}

For samples with a 350nm SRO buffer layer and varying STO
thickness, T$_C$ increases monotonically with in-plane strain as
shown in Fig.~\ref{fig:fig2.eps}(a), and in more detail in
Fig.~\ref{fig:fig2.eps}(b). The difference, $\Delta$T$_C$, is
about 14K (or $\Delta$T$_C$/T$_C$$\sim$10\%) between the 1000nm
STO sample and 50nm sample although the in-plane strain at room
temperature measured by glancing incidence x-ray diffraction (not
shown) only changes from 0.01\% to 0.23\%. The thinner film, with
greater strain, has the higher T$_C$, while thicker film has T$_C$
close to the bulk value. Thus the structural phase transition is
very sensitive to the biaxial strain caused by the lattice
mismatch at the interface, with the ratio
($\Delta$T$_C$/T$_C$)/$\varepsilon$ $\sim$45, where $\varepsilon$
is the in-plane strain.

We also studied a series of samples with the same 200nm STO layer
but varying the thickness of SRO buffer layer. Here we observed
even larger effect of strain on T$_C$, as shown in
Fig.~\ref{fig:fig2.eps}(c). In this case the largest strains and
the largest shifts in T$_C$ appear for SRO buffer layers of
intermediate thickness. We believe this is due to the complexity
of the SRO domain structure. The highest T$_C$ is in the sample
with a 20nm SRO buffer layer. In this case, T$_C$ is shifted about
30K above the STO film on a 350nm buffer and about 40K above the
most bulk like of our films [1000nm STO / 350nm SRO]. Clearly the
overall rise in T$_C$ can be quite large. Further detail will be
given in a follow up paper\cite{Hetb}, that also includes a
discussion of the temperature-strain phase diagram which has been
theoretically predicted\cite{Per00}.

\begin{figure}
\includegraphics[scale=.8]{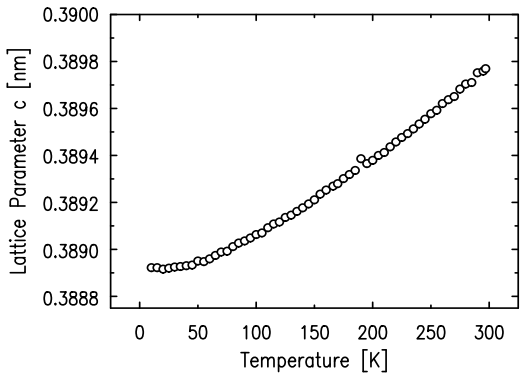}
\caption{\label{fig:fig3.eps}The temperature dependence of STO
lattice parameter \textit{c}, shows no abrupt change through the
whole range.}
\end{figure}

Although from the presence of the superlattice peaks we can see
that there is a phase transition near 130K, the out-of-plane
lattice constants evolve smoothly through the transition
temperature with no indication of the transition. This is
illustrated in Fig.~\ref{fig:fig3.eps} for the 200nm STO / 350nm
SRO film. The variation in lattice constant is merely the thermal
expansion with no indication of a phase transition. This is
obviously different from the bulk STO where a sudden slope change
is clearly seen at 105K\cite{Ale69}\cite{Hir95}. We ascribe this
effect to the clamping effect of the substrate upon the film. The
film in-plane lattice parameters are not free to move to their
bulk equilibrium values because they are tied to the substrate
lattice. Our experiments show that the overall film is tetragonal
at all temperatures measured. However this doesn't prevent the
TiO$_6$ octahedra from rotating and ordering leading to a reduced
symmetry phase. In fact the tetragonal cell shape seems to favor
the rotation of TiO$_6$ octahedra, stabilizing the lower symmetry
phase and enhancing T$_C$.

It is a remarkable that the phase transition in the STO films
involves a lowering of the internal symmetry due to TiO$_6$
octahedra rotations without a change in the unit cell size and
shape. This unique phase transition manifests a decoupling of the
internal degree of freedom of TiO$_6$ rotation and the external
degree of freedom of the unit cell dimensions. Such a difference
between the nature of structural phase transitions in epitaxial
films and bulk materials may occur in many situations and should
be considered in analyzing a variety of results.

Raman scattering experiments on similar films have studied the
temperature dependence of the structural R modes, which are Raman
active only below T$_C$\cite{Aki00}. In bulk STO these phonons
disappear at 105K, while in thin films their intensity vanishes at
about 120$\pm$5K. The mode is detectable in Raman upon the
doubling of the unit cell thus folding this mode back to the zone
center. This change is dependent upon the rotation of the
octahedra and not the presence of a tetragonal unit cell. In
contrast, dielectric measurements show obvious thickness
dependence and the temperature range over which the dielectric
constant is tunable by an electric field is vastly increased
versus the bulk\cite{Li98b}. The empirical evidence suggests that
this behavior is connected to the presence of a tetragonal unit
cell.

A review of the literature reveals that a similar effect may exist
in BaTiO$_3$ (BTO) films. Bulk BTO has a series of phase
transitions at 183K, 278K and 403K, at each of which the lattice
constants change drastically. All of the lower temperature phases
are ferroelectric. In contrast, lattice parameter measurements of
epitaxial films show no abrupt changes in either the \textit{a} or
\textit{c} constant, which both increase roughly linearly with
temperature from 15K to 800K\cite{Ter92}. However, it appears from
electrical measurements that the ferroelectric phase transition
still occurs in similar BTO films on a Pt/MgO
substrate\cite{Yon93}\cite{Yon01}. Ferroelectricity is due to the
shift of position of the Ti ion with respect to the O octahedron
surrounding it, not the cubic to tetragonal change in the unit
cell.  Thus ferroelectricity is due to the internal degree of
freedom of the phase transition. The ferroelectric transition in
BTO films seems to be another example of an epitaxy strain and
substrate clamping effect that permits a phase transition in the
internal atomic arrangements (ferroelectricity) but does not allow
external dimension changes.

It appears that the decoupling of the internal and external
aspects of a phase transition in films is not universal. One
example is NdNiO$_3$. Both bulk and films of this material have a
metal-insulator transition around 200K. Recent x-ray diffraction
measurements on epitaxial films of NdNiO$_3$ on LaAlO$_3$ have
shown that the out of plane lattice constant exhibits a sudden
change at the transition\cite{Heup}. Further experiments are
ongoing to try to determine what makes the two transitions
fundamentally different with respect to the clamping effect of the
substrate and the relaxation of the overall unit cell dimensions.

We acknowledge S. P. Alpay for helpful discussion. BW thanks the
A.P. Sloan Foundation Research Fellow program for partial support
of this work. Work at Brookhaven is supported by Division of
Material Sciences, U.S. Department of Energy under contract
DE-AC02-98CH10886. The work at Penn State is supported by NSF
under grant No. DMR-9702632, and work at UConn is supported by
the University of Connecticut Research Foundation.

\bibliography{sto}

\end{document}